# Correlation Analysis among Vorticity, Q method and Liutex

Yifei Yu[1], Pushpa Shrestha[1], Oscar Alvarez[1], Charles Nottage[1], Chaoqun Liu[1]*
1. Department of Mathematics, University of Texas at Arlington, Arlington, Texas 76019, USA
   cliu@exchange.uta.edu

Abstract:
Influenced by the fact that vorticity represents rotation for rigid body, people believe it also works for fluid flow. However, the theoretical predictions by vorticity do not match experiment results, which drove scientists to look for better methods to describe vortex. According to Dr. Liu's classification, all methods applied to detect vortex can be categorized into three generations. The vorticity-based method is classified as the first generation. Methods relying on eigenvalues of velocity gradient tensor are considered as the second generation. Although so many methods appeared, people still believe vorticity is vortex since vorticity theory looks perfect in math, and all other methods are only scalars and unable to indicate swirl direction. Recently, Dr. Liu innovated a new vortex identification method called Liutex. Liutex, a vector quantity, which is regarded as the third-generation method, not only overcomes all previous methods' drawbacks, but also has a clear physical meaning. The direction of Liutex represents the swirl axis of rotation, and its strength is equal to twice of angular speed. In this paper, we did a correlation analysis between vorticity, Q, $\lambda_{ci}$, $\lambda_2$ methods and Liutex based on a DNS case of boundary layer transition. The results show that the correlation between vorticity and Liutex is minimal in strong shear region, which demonstrates the idea that using vorticity to detect vortex lacks a scientific foundation; in other words, vorticity is not vortex.

Vorticity is an indicator of angular speed in solid mechanics. This fact has resulted in the assumption that vorticity could be used the same way in the field of fluid mechanics. In 1858, Helmholtz [1] suggested to use a vorticity tube/filament to display vortex structure. However, many experimental results show that vorticity does not match the actual results. The famous 2D Couette flow[2] is a type of laminar flow where there is no vortex. However, the vorticity is non-zero, conflicting with the non-rotational truth. A similar situation occurs in 2D laminar channel flow[3]. With the development of computer technology, Direct Numerical Analysis becomes doable in which the mesh scale is small enough to improve the accuracy. DNS research[4] has concluded that at some places, the rotation is strong; meanwhile, the magnitude of vorticity is minimal. The result of this evidence clearly indicates that vorticity is not vortex. This evidence drove researchers to find a better method to estimate vortex.

During the previous decades, several vortex identification methods have been proposed. The method that is most widely used is called Q criterion[5], which is dependent on the velocity gradient tensor. Another method, $\Delta$ criterion[6], regards the rotation area as where the velocity gradient tensor has complex eigenvalues. $\lambda_{ci}$ method[7], proposed is an extension of $\Delta$ criterion, uses the imaginary parts of the complex eigenvalues to measure the swirling strength. $\lambda_2$ method[8] defines the vortex area as the region where $S^2+W^2$ (where $S$ and $W$ are the symmetric and anti-symmetric parts of the velocity gradient tensor) has two negative eigenvalues. Although these methods perform better than vorticity, they share some common drawbacks. All of them are scalar, making iso-surface the only way to display the vortex structure. However, this causes the problem of having to select a proper threshold value. Also, because these methods are scalar, people are unable to locate the swirling axis. Dr. Liu[9] classified these eigenvalue-based methods as second-generation vortex identification methods and has formed a new physical concept—Liutex[10]. Liutex innovatively defines the local rotation axis, which works for both rigid body and fluid, solving the problem of locating the swirling axis and defined Liutex strength as the minimum of twice $\frac{\partial u}{\partial y}$, without shear contamination. After the occurrence of Liutex, many methods to do research in field of fluids have been developed and it gradually forms the Liutex system. Right now, this system includes Liutex similarity[11], Liutex core line[12,13], Liutex-Omega method[14,15] and etc. Recently, the Principal Coordinate and Principal Decomposition[16] were proposed. The Principal Coordinate is a unique coordinate under which rotation, shear, and stretching can be easily and correctly decomposed. Principal decomposition is the decomposition under Principal Coordinate.

Although many counterexamples and experiments can be provided to oppose the idea that vorticity is vortex, many text-books[17] still consider vorticity as the fundamental vortex identification method. So, in this letter, more evidence will be provided from a statistical perspective to convince people.

Correlation analysis, a data-based method, can be used to test the extent that two groups of data are related. If two variables are 100% relevant, then their correlation coefficient is 1. On the contrary, if they are 100% irrelevant, their correlation coefficient is zero.

In the following part of this letter, some previous methods and the recent Liutex method are reviewed.

***Definition 1***: $\omega$ is called vorticity if
$$\omega = \nabla \times v \tag{1}$$
$v$ is the velocity vector.

In solid mechanics, $\frac{1}{2}\omega$ is a good representation of rotation; however, this is not the case in fluid dynamics. In many recent papers[18], it has been shown that vorticity consists of rotation and shear. Where it is the so-called R-S decomposition of vorticity, namely, $\omega = R + S$, where R means rotation and S represents shear. So, $\frac{1}{2}\omega$ representing angular speed is only a special case in solid mechanics with the assumption that there is no shear.

***Definition 2***: Let $A$ and $B$ be the symmetric and anti-symmetric part of the velocity gradient tensor, then Q [5]is defined as
$$Q = \frac{1}{2}(\|B\|_F^2 - \|A\|_F^2) \tag{2}$$

where $\|\ \|_F$ means Frobenius norm.

The Q criterion performs better than vorticity and is widely used in the engineering field. It tries to remove the shear part from vorticity, as Q uses the symmetric part $A$ to approximate shear. However, $A$ is only an approximation of shear and not shear itself.

**Definition 3**: $\lambda_{ci}$ method[7] defines the strength of vortex as the imaginary part $\lambda_{ci}$ of the complex eigenvalue of the velocity gradient tensor $\nabla V$.

The $\lambda_{ci}$ method is derived from the idea that time-frozen streamlines display the flow rotation structure when the velocity gradient tensor $\nabla V$ has a pair of conjugate complex eigenvalues. In this situation, the tensor transformation of $\nabla V$ can be written as

$$\nabla V = \begin{bmatrix} v_r & v_{cr} & v_{ci} \end{bmatrix} \begin{bmatrix} \lambda_r & 0 & 0 \\ 0 & \lambda_{cr} & \lambda_{ci} \\ 0 & -\lambda_{ci} & \lambda_{cr} \end{bmatrix} \begin{bmatrix} v_r & v_{cr} & v_{ci} \end{bmatrix}^{-1} \quad (3)$$

Where $v_r$ is the eigenvector corresponding to the eigenvalue $\lambda_r$, and $v_{cr} \pm v_{ci}i$ are eigenvectors corresponding to the eigenvalues $\lambda_{cr} \pm \lambda_{ci}i$.

The equations of the streamlines can be expressed as

$$c_1(t) = c_1(0)e^{\lambda_r t} \quad (4)$$

$$c_2(t) = \left[c_2(0)\cos(\lambda_{ci}t) + c_3(0)\sin(\lambda_{ci}t)\right]e^{\lambda_{cr}t} \quad (5)$$

$$c_3(t) = \left[c_3(0)\cos(\lambda_{ci}t) - c_2(0)\sin(\lambda_{ci}t)\right]e^{\lambda_{cr}t} \quad (6)$$

where $\lambda_{ci}$ is used to indicate the rotation strength.

**Definition 4**: $\lambda_2$ method[8] defines the strength of vortex by using the second largest eigenvalue $\lambda_2$ of $A^2 + B^2$, where A and B are the symmetric and anti-symmetric parts of the velocity gradient tensor.

By assuming the fluid is non-viscous, the Navier-Stokes equation is converted to $A^2 + B^2 = -\nabla(\nabla p)/\rho$, where $p$ and $\rho$ represent pressure and density, respectively. Jeong and Hussain define the rotational area by the existence of two negative eigenvalues of the symmetric tensor $A^2 + B^2$.

**Definition 5**: The local rotation axis[10] is the normalized vector that satisfies:
$$d\boldsymbol{v} = \alpha d\boldsymbol{r} \quad (7)$$
$d\boldsymbol{v}$ is the increment of velocity, $d\boldsymbol{r}$ is the increment of $\boldsymbol{r}$, and $\alpha$ is a constant.

**Theorem 1**: The $\boldsymbol{r}$ in Def. 5 is the normalized eigenvector of velocity gradient tensor.[10]
Proof: Since $d\boldsymbol{v} = \nabla\boldsymbol{v} \cdot d\boldsymbol{r}$, then Def. 5 implies that $\nabla\boldsymbol{v} \cdot d\boldsymbol{r} = \alpha d\boldsymbol{r}$. Therefore, $d\boldsymbol{r}$ is the eigenvector of $\nabla v$.

**Definition 6**: Liutex is a vector method defined as[10]
$$\mathbf{R} = R\mathbf{r} \tag{8}$$
where $R$ is the magnitude of Liutex and $\mathbf{r}$ is the local rotation axis.
Wang [19] proposed an explicit expression of $R$ i.e.
$$R = [\boldsymbol{\omega} \cdot \mathbf{r} - \sqrt{(\boldsymbol{\omega} \cdot \mathbf{r})^2 - 4\lambda_{ci}^2}] \tag{9}$$
where $\boldsymbol{\omega}$ is vorticity, $\mathbf{r}$ is the local axis, and $\lambda_{ci}$ is the imaginary part of $\nabla v's$ complex eigenvalue.

**Definition 7**: X is a random variable with sample size n, $E(X)$ is the expectation of X if:
$$E(X) = \frac{1}{n}\sum_{i=1}^{n} X_i \tag{10}$$
$X_i$ is the i-th data of $X$.

**Definition 8**: $X$ is a random variable with sample size n, $Var(X)$ is the variance of $X$ if:
$$Var(X) = E[(X - E(X))^2] \tag{11}$$

**Definition 9**: $X$ is a random variable with sample size n, $\sigma(X)$ is the standard deviation of $X$ if:
$$\sigma(X) = \sqrt{Var(X)} \tag{12}$$

**Definition 10**: $X$ and $Y$ are random variables with sample size n, $cov(X,Y)$ is the correlation coefficient of $X$ and $Y$ if:
$$cov(X,Y) = E[(X - E(X))(Y - E(Y))] \tag{13}$$

**Definition 11**: $X$ and $Y$ are random variables with sample size n, $\rho(X,Y)$ is the correlation coefficient of $X$ and $Y$ if:
$$\rho(X,Y) = \frac{cov(X,Y)}{\sigma(X)\sigma(Y)} \tag{14}$$

The correlation coefficient is a statistical concept revealing the extent that two groups of data are related. If these two groups of data are entirely related, their correlation coefficient is 1. On the other hand, if they are entirely irrelevant, their correlation coefficient should be zero. The correlation coefficient provides a tool for us to test how closely related vorticity and Q method are to Liutex (Liutex correctly represents vortex) from a perspective of statistics.

Suppose the velocity gradient tensor under Principal Coordinate at the i-th time point is
$$\nabla V_i = \begin{bmatrix} \lambda_{cr-i} & -\frac{1}{2}R_i & 0 \\ \frac{1}{2}R_i + \varepsilon_i & \lambda_{cr-i} & 0 \\ \xi_i & \eta_i & \lambda_{r-i} \end{bmatrix} \tag{15}$$

The expression of vorticity is
$$\nabla \times V_i = (\eta_i, -\xi_i, R_i + \varepsilon_i) \tag{16}$$

The magnitude of vorticity can be written as

$$|\nabla \times V_i| = \sqrt{\eta_i^2 + \xi_i^2 + (R_i + \varepsilon_i)^2} \tag{17}$$

The expectation of vorticity can be expressed as

$$E(|\nabla \times V_i|) = \frac{1}{n}\sum_{j=1}^{n}\sqrt{\eta_j^2 + \xi_j^2 + (R_j + \varepsilon_j)^2} \tag{18}$$

The correlation between vorticity and Liutex is

$$\rho(|\nabla \times V_i|, R) = \frac{\sum_{i=1}^{n}\left[|\nabla \times V_i| - E(|\nabla \times V|)\right]\left(R_i - \frac{1}{n}\sum_{j=1}^{n}R_j\right)}{\sqrt{\sum_{i=1}^{n}\left[|\nabla \times V_i| - E(|\nabla \times V|)\right]^2}\sqrt{\sum_{i=1}^{n}\left(R_i - \frac{1}{n}\sum_{j=1}^{n}R_j\right)^2}} \tag{19}$$

Apply Cauchy-Stokes Decomposition on (16),

$$\nabla V_i = \begin{bmatrix} \lambda_{cr-i} & \frac{1}{2}\varepsilon_i & \frac{1}{2}\xi_i \\ \frac{1}{2}\varepsilon_i & \lambda_{cr-i} & \frac{1}{2}\eta_i \\ \frac{1}{2}\xi_i & \frac{1}{2}\eta_i & \lambda_{r-i} \end{bmatrix} + \begin{bmatrix} 0 & -\frac{1}{2}R_i - \frac{1}{2}\varepsilon_i & -\frac{1}{2}\xi_i \\ \frac{1}{2}R_i + \frac{1}{2}\varepsilon_i & 0 & -\frac{1}{2}\eta_i \\ \frac{1}{2}\xi_i & \frac{1}{2}\eta_i & 0 \end{bmatrix} = A + B \tag{20}$$

$$\begin{aligned} Q_i &= \frac{1}{2}\left(\|B_i\|_F^2 - \|A_i\|_F^2\right) \\ &= \frac{1}{2}\left[2\left(\frac{R_i}{2} + \frac{\varepsilon_i}{2}\right)^2 + 2\left(\frac{\xi_i}{2}\right)^2 + 2\left(\frac{\eta_i}{2}\right)^2\right] - \frac{1}{2}\left[2\lambda_{cr-i}^2 + \lambda_{r-i}^2 + 2\left(\frac{\varepsilon_i}{2}\right)^2 + 2\left(\frac{\xi_i}{2}\right)^2 + 2\left(\frac{\eta_i}{2}\right)^2\right] \\ &= \left(\frac{R_i}{2}\right)^2 + \frac{R_i\varepsilon_i}{2} - \lambda_{cr-i}^2 - \frac{\lambda_{r-i}^2}{2} \end{aligned} \tag{21}$$

The expectation of Q can be expressed as

$$E(Q) = \frac{1}{n}\sum_{i=1}^{n}\left[\left(\frac{R_i}{2}\right)^2 + \frac{R_i\varepsilon_i}{2} - \lambda_{cr-i}^2 - \frac{\lambda_{r-i}^2}{2}\right] \tag{22}$$

The correlation between Q and Liutex is in the form of

$$\rho(Q, R) = \frac{\sum_{i=1}^{n}\left[\left(\frac{R_i}{2}\right)^2 + \frac{R_i\varepsilon_i}{2} - \lambda_{cr-i}^2 - \frac{\lambda_{r-i}^2}{2} - E(Q)\right]\left(R_i - \frac{1}{n}\sum_{j=1}^{n}R_j\right)}{\sqrt{\sum_{i=1}^{n}\left[\left(\frac{R_i}{2}\right)^2 + \frac{R_i\varepsilon_i}{2} - \lambda_{cr-i}^2 - \frac{\lambda_{r-i}^2}{2} - E(Q)\right]^2}\sqrt{\sum_{i=1}^{n}\left(R_i - \frac{1}{n}\sum_{j=1}^{n}R_j\right)^2}} \tag{23}$$

The characteristic equation of (16) can be written as

$$(\lambda - \lambda_{r-i})\left[(\lambda - \lambda_{cr-i})^2 + \frac{R_i}{(R_i + 2\varepsilon_i)}\right] = 0 \tag{24}$$

So, the three eigenvalues are

$$\lambda_1 = \lambda_{r-i} \tag{25}$$

$$\lambda_2 = \lambda_{cr-i} + i\sqrt{\frac{R_i}{2}\left(\frac{R_i}{2} + \varepsilon_i\right)} \tag{26}$$

$$\lambda_2 = \lambda_{cr-i} - i\sqrt{\frac{R_i}{2}\left(\frac{R_i}{2} + \varepsilon_i\right)} \tag{27}$$

Since the rotation matrix is orthogonal, the eigenvalues of velocity gradient tensor do not change.

$$\lambda_2 = \lambda_{cr-i} + i\sqrt{\frac{R_i}{2}\left(\frac{R_i}{2} + \varepsilon_i\right)} = \lambda_{cr-i} + i\lambda_{ci-i} \tag{28}$$

Thus,

$$\lambda_{ci-i} = \sqrt{\frac{R_i}{2}\left(\frac{R_i}{2} + \varepsilon_i\right)} \tag{29}$$

The expectation of $\lambda_{ci}$ is

$$E(\lambda_{ci}) = \frac{1}{n}\sum_{i=1}^{n}\sqrt{\frac{R_i}{2}\left(\frac{R_i}{2} + \varepsilon_i\right)} \tag{30}$$

Correlation between $\lambda_{ci}$ and Liutex is

$$\rho(\lambda_{ci}, R) = \frac{\sum_{i=1}^{n}\left[\sqrt{\frac{R_i}{2}\left(\frac{R_i}{2} + \varepsilon_i\right)} - E(\lambda_{ci})\right]\left(R_i - \frac{1}{n}\sum_{j=1}^{n}R_j\right)}{\sqrt{\sum_{i=1}^{n}\left[\sqrt{\frac{R_i}{2}\left(\frac{R_i}{2} + \varepsilon_i\right)} - E(\lambda_{ci})\right]^2}\sqrt{\sum_{i=1}^{n}\left[R_i - \frac{1}{n}\sum_{j=1}^{n}R_j\right]^2}} \tag{31}$$

If no stretching or shear exists, $\lambda_{ci-i} = \frac{R_i}{2}$, in which case, $\rho(\lambda_{ci}, R) = 1$.

In a DNS research of boundary layer transition[20], the grid level is $1920 \times 128 \times 241$, representing the total amount in streamwise(x), spanwise(y), and wall-normal(z) directions. The first interval length in normal direction at the origin is 0.43 in wall units ($Z^+ = 0.43$). The flow parameters, including Mach number, Reynolds number, etc.… are listed in table 1. Here, $x_{in} = 300.79\delta_{in}$ represents the distance between the leading edge and the inlet, Lx, Ly, and Lz$_{in}$ are the lengths of the computational domain and $T_w$ is the wall temperature.

| $M_\infty$ | Re | $x_{in}$ | Lx | Ly | Lz$_{in}$ | $T_w$ | $T_\infty$ |
|---|---|---|---|---|---|---|---|
| 0.5 | 1000 | $300.79\delta_{in}$ | $798.03\delta_{in}$ | $22\delta_{in}$ | $40\delta_{in}$ | 273.15K | 273.15K |

Table 1  Flow parameters

Three different x positions, 402.8, 500.7, and 815.5, which are in the area of laminar, transitional, and turbulent flow, respectively, are selected. For each x position, ten different z positions are chosen, which can be found in Table 2. 400 snapshots are used to calculate the correlation coefficient.

The positions of z are shown in the following table.

| $z_1$ | $z_2$ | $z_3$ | $z_4$ | $z_5$ | $z_6$ | $z_7$ | $z_8$ | $z_9$ | $z_{10}$ |
|---|---|---|---|---|---|---|---|---|---|
| 0.26 | 0.38 | 0.50 | 0.65 | 0.81 | 1.08 | 1.35 | 1.70 | 2.17 | 2.77 |

Table 2 selected points' z positions

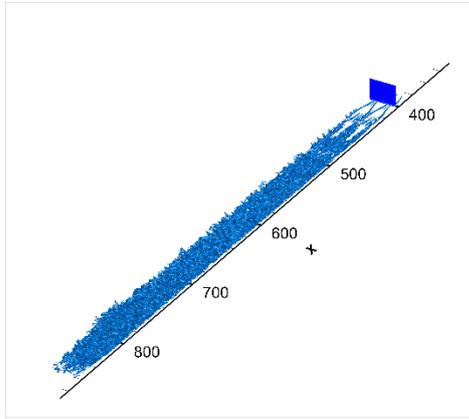 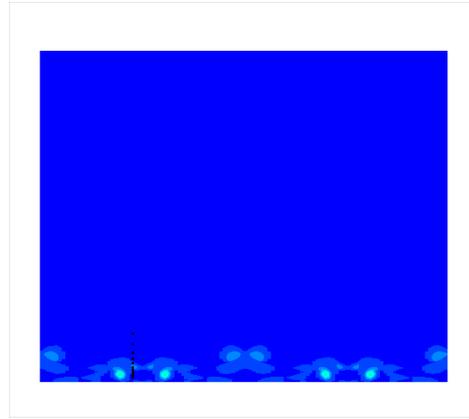

figure 1　　Points(black) selected at x=402.8

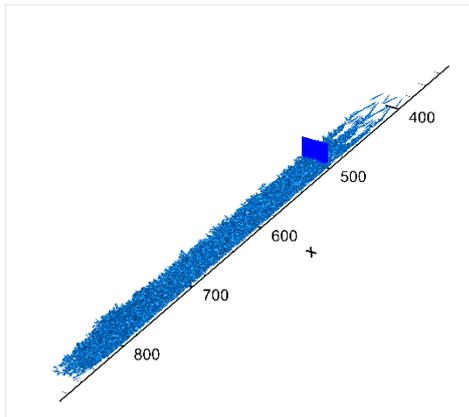 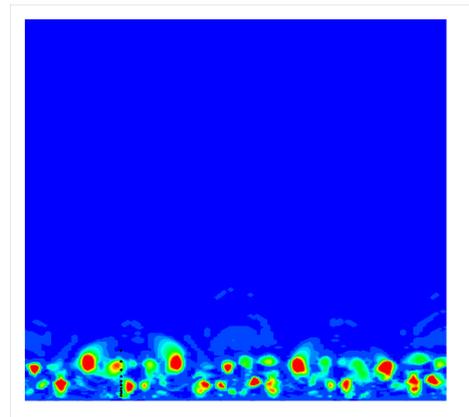

figure 2　　Points(black) selected at x=500.7

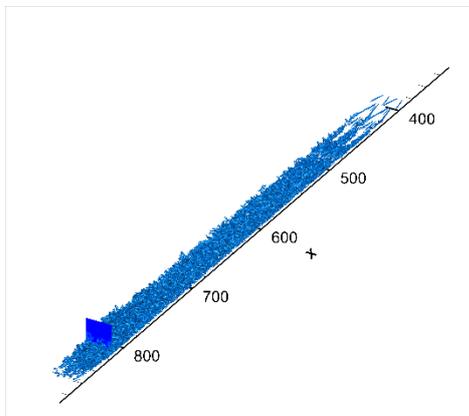 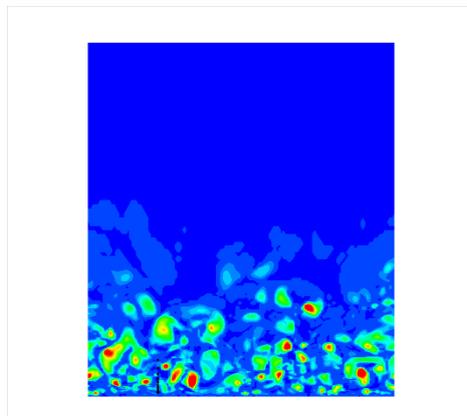

figure 3　　Points(black) selected at x=815.5

The correlation coefficients for x=402.8 and y=4.98 are shown in table 3 below.

|  | $z_1$ | $z_2$ | $z_3$ | $z_4$ | $z_5$ | $z_6$ | $z_7$ | $z_8$ | $z_9$ | $z_{10}$ |
|---|---|---|---|---|---|---|---|---|---|---|
| $\rho(R,Vor)$ | -0.59 | -0.61 | -0.24 | 0.43 | -0.26 | -0.51 | -0.62 | -0.43 | 0.89 | 0.93 |
| $\rho(R,Q)$ | 0.71 | 0.63 | 0.54 | 0.70 | 0.56 | 0.92 | 0.84 | 0.78 | 0.80 | 0.94 |
| $\rho(R,\lambda_{ci})$ | 0.95 | 0.94 | 0.92 | 0.88 | 0.81 | 0.94 | 0.89 | 0.84 | 0.81 | 0.97 |
| $\rho(R,-\lambda_2)$ | 0.80 | 0.73 | 0.47 | 0.60 | 0.66 | 0.93 | 0.90 | 0.76 | 0.85 | 0.95 |

Table 3 Correlation between Liutex and vorticity, Q, $\lambda_{ci}$ and $\lambda_2$ at x=402.8

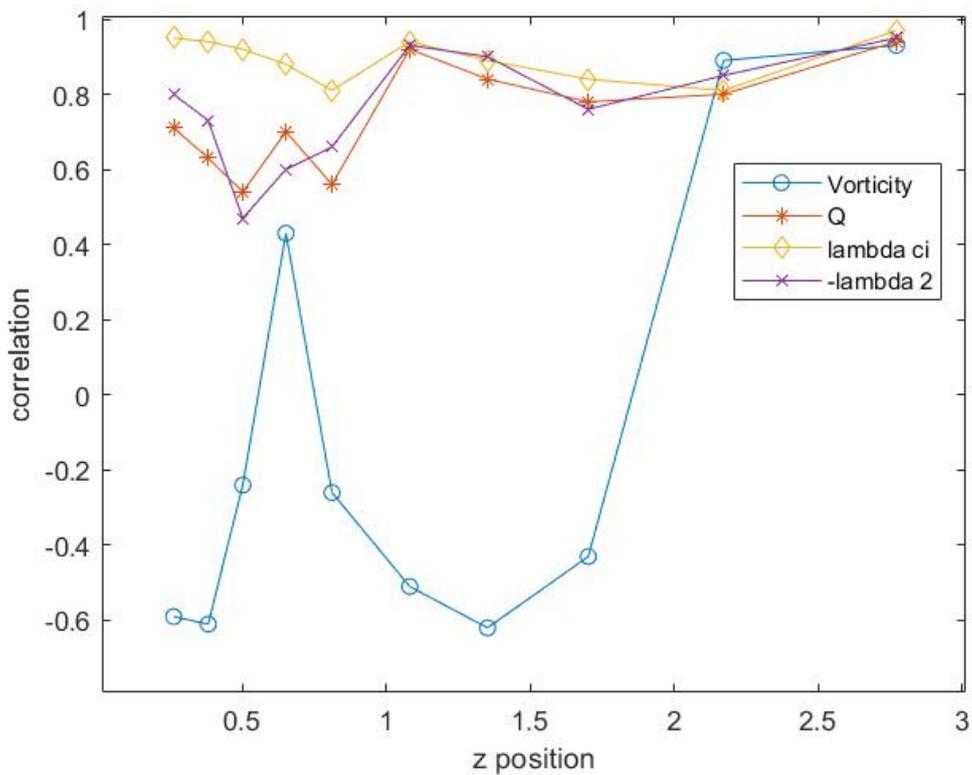

figure 4 Correlation between Liutex and vorticity, Q, $\lambda_{ci}$ and $\lambda_2$ at x=402.8

The correlation coefficients for x=500.7 and y=4.98 are shown in table 4 below.

|  | $z_1$ | $z_2$ | $z_3$ | $z_4$ | $z_5$ | $z_6$ | $z_7$ | $z_8$ | $z_9$ | $z_{10}$ |
|---|---|---|---|---|---|---|---|---|---|---|
| $\rho(R,Vor)$ | -0.05 | 0.007 | -0.06 | 0.26 | 0.42 | 0.58 | 0.56 | 0.57 | 0.83 | 0.79 |
| $\rho(R,Q)$ | 0.77 | 0.73 | 0.72 | 0.83 | 0.32 | 0.86 | 0.85 | 0.81 | 0.92 | 0.88 |
| $\rho(R,\lambda_{ci})$ | 0.91 | 0.87 | 0.86 | 0.91 | 0.92 | 0.93 | 0.95 | 0.93 | 0.98 | 0.96 |
| $\rho(R,-\lambda_2)$ | 0.81 | 0.80 | 0.82 | 0.86 | 0.63 | 0.91 | 0.89 | 0.85 | 0.92 | 0.90 |

Table 4 Correlation between Liutex and vorticity, Q, $\lambda_{ci}$ and $\lambda_2$ at x=500.7

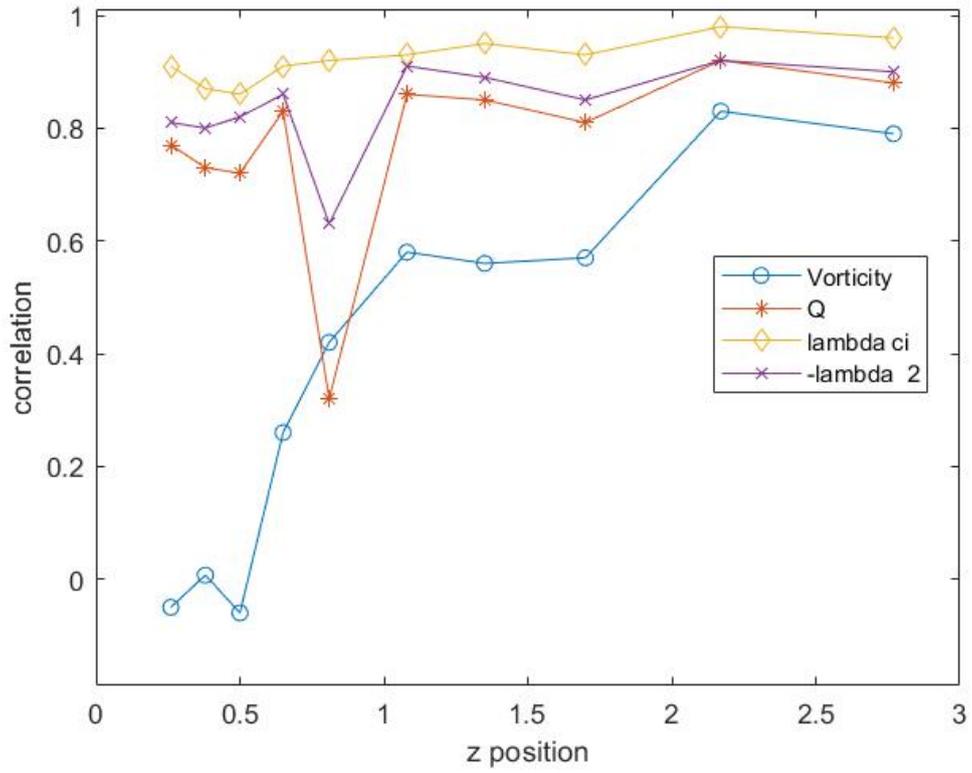

figure 5 Correlation between Liutex and vorticity, Q, $\lambda_{ci}$ and $\lambda_2$ at x=500.7

The correlation coefficients for x=815.5 and y=4.98 are shown in table 5 below.

|  | $z_1$ | $z_2$ | $z_3$ | $z_4$ | $z_5$ | $z_6$ | $z_7$ | $z_8$ | $z_9$ | $z_{10}$ |
|---|---|---|---|---|---|---|---|---|---|---|
| $\rho(R,Vor)$ | 0.10 | 0.03 | 0.08 | 0.30 | 0.55 | 0.73 | 0.41 | 0.81 | 0.71 | 0.47 |
| $\rho(R,Q)$ | 0.90 | 0.86 | 0.70 | 0.69 | 0.83 | 0.79 | 0.69 | 0.85 | 0.84 | 0.68 |
| $\rho(R,\lambda_{ci})$ | 0.91 | 0.90 | 0.92 | 0.88 | 0.90 | 0.94 | 0.90 | 0.96 | 0.95 | 0.93 |
| $\rho(R,-\lambda_2)$ | 0.90 | 0.87 | 0.80 | 0.75 | 0.89 | 0.88 | 0.78 | 0.89 | 0.87 | 0.74 |

Table 5 Correlation between Liutex and vorticity, Q, $\lambda_{ci}$ and $\lambda_2$ at x=815.5

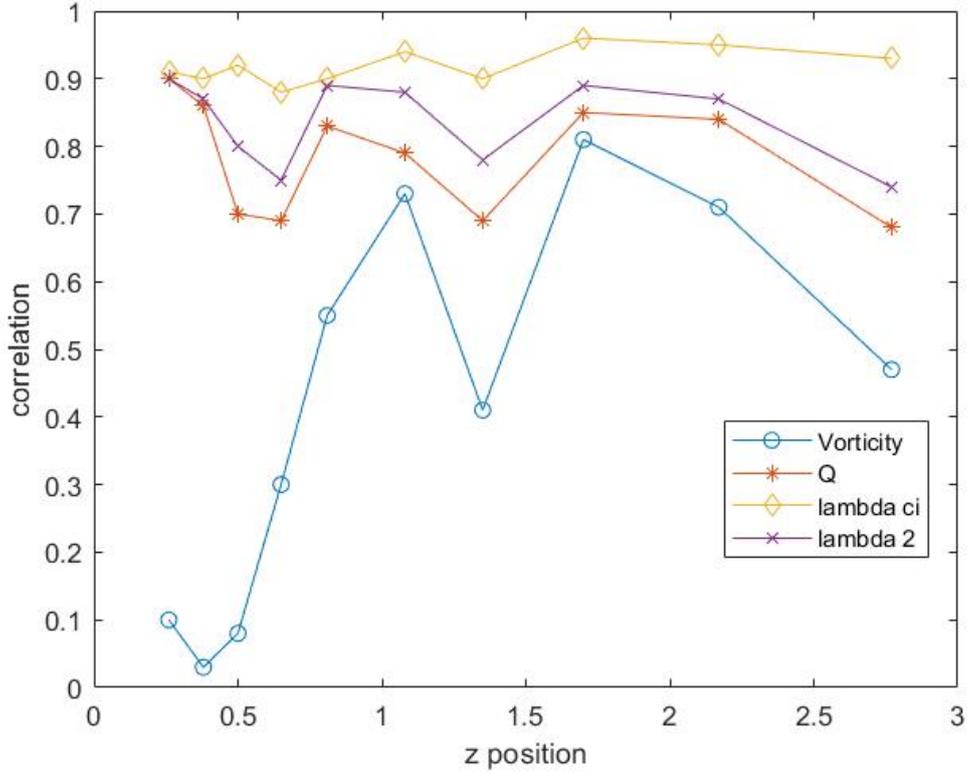

figure 6 Correlation between Liutex and vorticity, Q, $\lambda_{ci}$ and $\lambda_2$ at x=815.5

Generally, vorticity is the worst method to detect vortex, as seen in each figure, the line of correlation between Liutex and vorticity is always at the bottom. The reason for this phenomenon can be founded from (18), $|\nabla \times V_i| = \sqrt{\eta_i^2 + \xi_i^2 + (R_i + \varepsilon_i)^2}$, vorticity is contaminated by shear from three different directions, while the other methods are only contaminated by shear from one or two directions. An apparent tendency that can be observed from the figures is, with the increase of distance from the boundary, vorticity's correlation goes up, which corresponds with the truth that shears are strong near boundary region and weak away from the boundary. It also shows the correctness of (18).

Although Q and $\lambda_2$ methods perform better than vorticity, these two methods share one significant problem that is their dimensions are not correct. From (24) $Q_i = \left(\dfrac{R_i}{2}\right)^2 + \dfrac{R_i \varepsilon_i}{2} - \lambda_{cr-i}^2 - \dfrac{\lambda_{r-i}^2}{2}$, it can be seen that Q is roughly related to $R^2$, which is the square of the angular speed, rather than exactly rotation speed. So even though the Q or $\lambda_2$ method can indicate the relative strength of rotation, the numbers given by these two methods lack correct physical meaning.

An interesting phenomenon is that the correlation between vorticity and Liutex could be a negative number, which seems to conflict with R-S decomposition $\omega = R + S$. However, indeed, it does not. The key point is, $\omega = R + S$ is a vector relation, but the correlation is based on scalar, so it is possible that vorticity magnitude increases when R magnitude decreases. In our numerical simulation, there is no vortex in the origin, since it is laminar flow; however, it does have vorticity. Also, gradually, vortex occurs, leading to the increase of Liutex, while vorticity may decrease because the shear part of vorticity is converted to vortex. This is another proof that vorticity is not vortex.

In summary, this letter explains from a correlation analysis perspective, both theoretically and numerically, that vorticity and all the other previous vortex identification methods cannot represent vortex properly.

1. In any theoretical session, it is shown that previous vortex identification methods are contaminated by shear and stretching, which can be observed as well in numerical correlation analysis in the regions where shear and stretching are strong.
2. Vorticity, which has the worst performance among previous methods, is apparently not a good indicator of vortex. R-S decomposition of vorticity clearly reveals that vorticity consists of rotation and shear rather than pure rotation. This demonstrates the reason that vorticity works well for rigid bodies, for which it is assumed there is no shear but does not perform well for fluids, for which shear cannot be omitted. Also, the correlation between vorticity and Liutex can be negative. Therefore, it is of considerable importance to correct the idea that vorticity is vortex, especially in the content of textbooks.
3. Generally, $\rho(\lambda_{ci}, R) \approx \rho(\lambda_2, R) \approx \rho(Q, R) > \rho(Vorticity, R)$. The previous vortex identification methods can improve vortex detecting accuracy to some extent, and that is why they are widely used in engineering fields that are related to fluids. However, these methods still do not represent exact rotation. The dimensions of Q and $\lambda_2$ methods are roughly square of angular speed resulting in the ambiguity of their physical meanings. Liutex extracts the rigid rotation part from its motion, so it makes more sense to use Liutex over these previous methods that only try to represent vortex.

Reference:
[1] Von Herrn H. Helmholtz, "über Integrale der hydrodynamischen Gleichungen, welche den Wirbelbewegungen Entsprechen," Journal für die reine und angewandte Mathematik **1858(55)**, 25 (1858).
[2] N. Tillmark and P. H. Alfredsson, "Experiments on transition in plane Couette flow," J. Fluid Mech. **235** (-1), 89 (1992).
[3] S. P. Sutera and R. Skalak, "The History of Poiseuille's Law," Annu. Rev. Fluid Mech. **25** (1), 1 (1993).
[4] Y. Wang, Y. Yang, G. Yang, and C. Liu, "DNS Study on Vortex and Vorticity in Late Boundary Layer Transition," Commun. Comput. Phys. **22** (2), 441 (2017).
[5] J.C.R. Hunt, A.A. Wray, and P.Moin, "Eddies, stream, and convergence zones in turbulent

flows," Center for turbulence research report CTR-S88, 193 (1988).

[6] M. S. Chong, A. E. Perry, and B. J. Cantwell, "A general classification of three‐dimensional flow fields," Physics of Fluids A: Fluid Dynamics **2** (5), 765 (1990).

[7] J. ZHOU, R. J. ADRIAN, S. BALACHANDAR, T. M. KENDALL, "Mechanisms for generating coherent packets of hairpin vortices in channel flow," Journal of fluid mechanics **387**, 353 (1999).

[8] J. Jeong and F. Hussain, "On the identification of a vortex," J. Fluid Mech. **285** (-1), 69 (1995).

[9] C. Liu, Y.-s. Gao, X.-r. Dong, Y.-q. Wang, J.-m. Liu, Y.-n. Zhang, X.-s. Cai, and N. Gui, "Third generation of vortex identification methods: Omega and Liutex/Rortex based systems," J Hydrodyn **31** (2), 205 (2019).

[10] C. Liu, Y. Gao, S. Tian, and X. Dong, "Rortex—A new vortex vector definition and vorticity tensor and vector decompositions," Physics of Fluids **30** (3), 35103 (2018).

[11] W.-q. Xu, Y.-q. Wang, Y.-s. Gao, J.-m. Liu, H.-s. Dou, and C. Liu, "Liutex similarity in turbulent boundary layer," J Hydrodyn **31** (6), 1259 (2019).

[12] Y.-s. Gao, J.-m. Liu, Y.-f. Yu, and C. Liu, "A Liutex based definition and identification of vortex core center lines," J Hydrodyn **31** (3), 445 (2019).

[13] H. Xu, X.-s. Cai, and C. Liu, "Liutex (vortex) core definition and automatic identification for turbulence vortex structures," J Hydrodyn **31** (5), 857 (2019).

[14] X. Dong, Y. Gao, and C. Liu, "New normalized Rortex/vortex identification method," Physics of Fluids **31** (1), 11701 (2019).

[15] J. Liu and C. Liu, "Modified normalized Rortex/vortex identification method," Physics of Fluids **31** (6), 61704 (2019).

[16] Y. Yu, P. Shrestha, C. Nottage, and C. Liu, "Principal coordinates and principal velocity gradient tensor decomposition," J Hydrodyn **32** (3), 441 (2020).

[17] C. K. Batchelor, *An introduction to fluid dynamics* (Cambridge University Press, Cambridge, 2010).

[18] Y. Gao, Y. Yu, J. Liu, and C. Liu, "Explicit expressions for Rortex tensor and velocity gradient tensor decomposition," Physics of Fluids **31** (8), 81704 (2019).

[19] Y.-q. Wang, Y.-s. Gao, J.-m. Liu, and C. Liu, "Explicit formula for the Liutex vector and physical meaning of vorticity based on the Liutex-Shear decomposition," J Hydrodyn **31** (3), 464 (2019).

[20] C. Liu, Y. Yan, and P. Lu, "Physics of turbulence generation and sustenance in a boundary layer," Computers & Fluids **102**, 353 (2014).